\pdfoutput=1
\documentclass[11pt,twoside,a4paper,cmspaper,final,collab]{cms-tdr}
\def\svnVersion{f9f2193-D}\def\svnDate{2021/06/10}\def\cmsCernNoTag{CERN-EP-2020-197}\def\cmsCernDate{\today}\def\cmsMessage{"Published in Physics Letters B as \href{http://dx.doi.org/10.1016/j.physletb.2021.136409}{\doi{10.1016/j.physletb.2021.136409}.}"}
\begin{document}\cmsNoteHeader{SMP-20-008}

\ifthenelse{\boolean{cms@external}}{\providecommand{\cmsLeft}{upper\xspace}}{\providecommand{\cmsLeft}{left\xspace}}
\ifthenelse{\boolean{cms@external}}{\providecommand{\cmsRight}{lower\xspace}}{\providecommand{\cmsRight}{right\xspace}}
\providecommand{\CL}{CL\xspace}
\newcommand{\ptW}{\ensuremath{\pt^{\PW}}\xspace}
\newcommand{\eff}{\ensuremath{\varepsilon_{\ell}}\xspace}
\newcommand{\mpg}{\ensuremath{m_{\Pgp\Pgg}}\xspace}

\cmsNoteHeader{SMP-20-008}

\title{Search for the rare decay of the \texorpdfstring{\PW}{W} boson into a pion and a photon in proton-proton collisions at \texorpdfstring{$\sqrt{s} = 13\TeV$}{sqrt(s) = 13 TeV}}

\date{\today}

\abstract{A search is performed for the rare decay $\PWpm\to\Pgppm\Pgg$ in proton-proton collisions at $\sqrt{s} = 13\TeV$. Data corresponding to an integrated luminosity of $137\fbinv$ were collected during 2016 to 2018 with the CMS detector. This analysis exploits a novel search strategy based on {\PW} boson production in top quark pair events. An inclusive search for the $\PWpm\to\Pgppm\Pgg$ decay is not optimal at the LHC because of the high trigger thresholds. Instead, a trigger selection is exploited in which the {\PW} boson originating from one of the top quarks is used to tag the event in a leptonic decay. The {\PW} boson emerging from the other top quark is used to search for the $\PWpm\to\Pgppm\Pgg$ signature. Such decays are characterized by an isolated track pointing to a large energy deposit, and by an isolated photon of large transverse momentum. The presence of b quark jets reduces the background from the hadronization of light-flavor quarks and gluons. The $\PWpm\to\Pgppm\Pgg$ decay is not observed. An upper exclusion limit is set to this branching fraction, corresponding to $1.50\times 10^{-5}$ at $95\%$ confidence level, whereas the expected upper exclusion limit is $0.85^{+0.52}_{-0.29} \times 10^{-5}$.}

\hypersetup{%
pdfauthor={CMS Collaboration},%
pdftitle={Search for the rare decay of the W boson into a pion and a photon in proton-proton collisions at sqrt(s) = 13 TeV},%
pdfsubject={CMS},%
pdfkeywords={CMS, standard model, EWK}}

\maketitle

\section{Introduction}
Rare hadronic decays of {\PW} bosons represent probes of the strong interaction at the boundary between the perturbative and nonperturbative domains of quantum chromodynamics (QCD), offering insights into factorization and meson form factors at large energy scales \cite{bib:w_z_radiative,bib:w_exc_had_decays,bib:H_to_light_quarks}. In particular, exclusive decays of the {\PW} boson into final states containing a single meson can be used to test these theoretical frameworks in a context where the energy scales are sufficiently large. At such scales, corrections that depend on the momentum transferred to the hadron in the final state can be neglected. The observation of these decays would validate the QCD factorization formalism, and enhance the possibility to perform precise calculations using such an approach. Furthermore, at future colliders they could provide a new way to measure the mass of the {\PW} boson that is based solely on visible single-particle decay products. Theoretical calculations for such branching fractions ($\mathcal{B}$) have large uncertainties: the expected value of $\mathcal{B}(\PWpm\to\Pgppm\Pgg)$, for instance, is in the range of $10^{-9}-10^{-7}$ \cite{bib:w_z_radiative,bib:art_mangano}. Upper limits on $\mathcal{B}(\PWpm\to\Pgppm\Pgg)$ and on two other exclusive {\PW} boson decays were set previously at $95\%$ confidence level (\CL): $\mathcal{B}(\PWpm\to \Pgppm\Pgg) < 7.0\times10^{-6}$ \cite{bib:cdf_wpigamma}, $\mathcal{B}(\PWpm\to\PDpms\Pgg) < 1.3\times10^{-3}$ \cite{bib:cdf_wdgamma}, and $\mathcal{B}(\PWpm\to\Pgppm\Pgppm\PGpmp) < 1.01\times10^{-6}$ \cite{bib:cms_w3pi}. For the {\PZ} boson, there are comparatively many more results on searches for such rare exclusive decay channels \cite{bib:PDG}.

This Letter reports the results of the first search at the CERN LHC for the rare decay $\PWpm\to\Pgppm\Pgg$. The high energy thresholds for single photon triggers make an inclusive search, similar to the one performed by the CDF Collaboration \cite{bib:cdf_wpigamma}, unsuitable at CMS. Therefore, this analysis uses a novel approach \cite{bib:art_mangano}, focusing on {\PW} boson production from top quark-antiquark pair (\ttbar) events at the center-of-mass energy of $13\TeV$. The muon or electron emerging from the leptonic decay of one of the two {\PW} bosons from the \ttbar pair is used to select events at the trigger level. The other {\PW} boson in the event is employed to search for the rare decay $\PWpm\to\Pgppm\Pgg$. The decay is characterized by an isolated photon of large transverse momentum (\pt), and an isolated track corresponding to a particle with large \pt. The analysis exploits the reduced jet activity expected in the vicinity of the pion \cite{bib:art_mangano}, which is produced via electroweak processes, to reject background pions. The {\cPqb} quark jets are used in the multivariate event selection to reduce the contribution from non \ttbar backgrounds. The $\PWpm\to\Pgppm\Pgg$ branching fraction is determined by parametrizing the number of observed signal events ($N_{\text{sig}}$) as:
\begin{linenomath}
\ifthenelse{\boolean{cms@external}}
{\begin{multline}
N_{\text{sig}} = \sigma_{\ttbar}\,\mathcal{L}_{\text{int}}\,\mathcal{B}(\PWpm\to\Pgppm\Pgg)\,\sum_{\ell}\mathcal{B}(\PWmp\to\ell^{\mp }\nu)\,\eff,\\
\ell=\Pgm,\Pe,\Pgt_{\ell},
\end{multline}\label{eq:Nsig}}
{\begin{equation}\label{eq:Nsig}
N_{\text{sig}} = \sigma_{\ttbar}\,\mathcal{L}_{\text{int}}\,\mathcal{B}(\PWpm\to\Pgppm\Pgg)\,\sum_{\ell}\mathcal{B}(\PWmp\to\ell^{\mp }\nu)\,\eff,~~~~\ell=\Pgm,\Pe,\Pgt_{\ell},
\end{equation}}
\end{linenomath}
where $\sigma_{\ttbar}$ is the \ttbar production cross section, $\mathcal{B}(\PWmp\to\ell^{\mp }\nu)$ the branching fraction of the {\PW} boson into leptons, $\mathcal{L}_{\text{int}}$ the integrated luminosity, and \eff the efficiency for the whole selection procedure (including acceptance effects). The symbol $\Pgt_{\ell}$ indicates that the leptonic decays of the $\Pgt$ lepton originating from $\PW\to\Pgt\nu$ are also considered in the computation.

\section{The CMS detector}
The central feature of the CMS apparatus is a superconducting solenoid 6\unit{m} in internal diameter, providing a magnetic field of 3.8\unit{T}. A silicon pixel and strip tracker, a lead tungstate crystal electromagnetic calorimeter (ECAL), and a brass and scintillator hadron calorimeter reside within the solenoid volume. Each of these systems is composed of a barrel and two endcap sections. Forward hadron calorimeters extend the pseudorapidity ($\eta$) coverage provided by the barrel and endcap detectors. Muons are detected in gas-ionization chambers embedded in the steel flux-return yoke outside the solenoid.

Events of interest are selected using a two-tiered trigger system \cite{bib:run1_trigger}. The first level, composed of custom hardware processors, uses information from the calorimeters and muon detectors to select events at a rate of around 100\unit{kHz} with a latency of about 4\mus. The second level, known as the high-level trigger, consists of a farm of processors running a version of the full event reconstruction software optimized for fast processing, and reduces the event rate to around 1\unit{kHz} before data storage. A more detailed description of the CMS detector, together with a definition of its coordinate system and relevant kinematic variables, can be found in Ref. \cite{bib:cms}.

\section{Data samples and simulation}\label{sec:samples}
The analyzed collision data correspond to an integrated luminosity of $137\fbinv$ recorded by the CMS experiment in proton-proton ($\Pp\Pp$) collisions at $\sqrt{s}=13\TeV$ during 2016 ($35.9\fbinv$), 2017 ($41.5\fbinv$), and 2018 ($59.7\fbinv$).

The signal process is simulated at leading order (LO) in perturbative QCD using the Monte Carlo (MC) event generator \PYTHIA 8.226 (8.230) for 2016 (2017 and 2018) \cite{cit:PYTHIA8}. The transverse momentum of the {\PW} boson ({\ptW}) decaying into a pion and a photon in the signal MC is weighted so that it matches that obtained in a next-to-LO (NLO) $\PW\to \ell\nu$ ($\ell=\Pgm, \Pe$) simulation in \POWHEG v2.0 \cite{cit:POWHEG_1,cit:POWHEG_2,cit:POWHEG_3,bib:ttbar_POWHEG}, where the {\PW} boson originates from \ttbar decay. This correction does not affect the total signal normalization.

The main background processes in this analysis are nonsignal \ttbar production, Drell--Yan events, associated production of a vector boson and a photon or jets ($\PW\Pgg\to\ell\nu\Pgg$, $\PZ\Pgg\to\ell\ell\Pgg$ and $\PW+\text{jets}\to\ell\nu+\text{jets}$), along with standard model events comprised uniquely of jets produced through the strong interaction, referred to as QCD multijet events. The \ttbar backgrounds are simulated at NLO using the \POWHEG v2.0 framework \cite{bib:ttbar_POWHEG}, whereas the Drell--Yan, $\PW\Pgg\to\ell\nu\Pgg$, $\PZ\Pgg\to\ell\ell\Pgg$, $\PW+\text{jets}\to\ell\nu+\text{jets}$, and QCD multijet processes are simulated at NLO using the \MGvATNLO 2.2.2 (2.4.2) generator for 2016 (2017 and 2018) \cite{cit:aMCatNLO}.

The NNPDF 3.0 NLO \cite{bib:NNPDF_3.0} (NNPDF 3.1 next-to-NLO \cite{bib:NNPDF_3.1}) parton distribution functions are used for generating all 2016 (2017 and 2018) MC samples. The generators are interfaced to \PYTHIA for parton showering and parton fragmentation. The \PYTHIA parameters affecting the description of the underlying event are set to the CUETP8M1 tune \cite{bib:CUETP8M1} (CP5 tune \cite{bib:CP5}) for the 2016 (2017 and 2018) simulation. For the 2016 \ttbar backgrounds, the CP5 tune is used instead of CUETP8M1. The difference in generator version and configuration over the years of data taking reflects the improved knowledge of the production mechanisms in $\Pp\Pp$ collisions at CMS.

All the simulated background samples are normalized to the cross sections obtained from the corresponding event generator. For the signal, the \ttbar cross section measured by CMS with the 2015 data \cite{bib:ttbar_xsec} is used, together with the most up-to-date measurements of $\mathcal{B}(\PW\to\ell\nu)$ \cite{bib:PDG}.

Generated events are processed through a simulation of the CMS detector based on \GEANTfour \cite{cit:GEANT4} and reconstructed using the same algorithms used for collision data. Additional $\Pp\Pp$ collisions in the same or nearby bunch crossings (pileup) are simulated and their distributions are weighted to reproduce the pileup multiplicity measured in data, which has an average of about 23 (32) interactions per bunch crossing in 2016 (2017 and 2018).

\section{Event reconstruction}
Events are reconstructed using the particle-flow (PF) algorithm \cite{cit:PF}, which combines information from the subdetectors to reconstruct and identify individual particles. The candidate $\Pp\Pp$ interaction vertex with the largest value of summed physics-object $\pt^2$ is referred to as the primary vertex (PV). The physics objects are the jets, clustered using the jet finding algorithm \cite{bib:anti_kt,bib:fastjet} with the tracks assigned to candidate vertices as inputs, and the associated missing transverse momentum \ptvecmiss, computed as the negative vector \pt sum of all the PF candidates in an event. This vector, whose magnitude is denoted as \ptmiss~\cite{bib:MET}, is modified to account for corrections to the energy scale of the reconstructed jets in the event.

Muon candidates are reconstructed by identifying signals from the muon subsystems together with those from the inner tracker \cite{cit:cms_muonSyst}. The inner detector track must be consistent with originating from the PV. The energy of muons is obtained from the curvature of the corresponding track. The \pt sum of the PF candidates within a cone of $\DR \equiv \sqrt{\smash[b]{(\Delta\eta)^2+(\Delta\phi)^2}}< 0.4$ around the muon direction is required to be $<$25\% of the muon \pt after implementing a correction for neutral PF objects not associated with the PV. Here $\phi$ indicates the azimuthal angle in radians in the plane transverse to the beam axis.

Electron candidates are reconstructed from clusters in the ECAL that are matched to tracks reconstructed using a Gaussian-sum filter algorithm. The track is required to be consistent with originating from the PV. Electron identification proceeds through a multivariate classifier that involves observables sensitive to bremsstrahlung along the electron trajectory, the geometrical and energy-momentum compatibility between the electron track and the associated energy cluster in the ECAL, the energy distribution of the electromagnetic shower, and discrimination against electrons originating from photon conversion. The PF-based isolation variables are also used as input variables in the multivariate classifier, including the \pt of photons, charged and neutral hadrons in the event \cite{cit:cms_electron_reco}.

Photon reconstruction is performed in the region $\abs{\eta} < 2.5$, excluding the ECAL barrel-endcap transition region $1.44 < \abs{\eta} < 1.57$. The energies of photons are obtained from the ECAL measurement. An electron veto is implemented to minimize photon misidentification, combining information from the ECAL and the pixel tracker. As in the electron case, photon identification proceeds through a multivariate classifier that makes use of PF-based isolation variables \cite{bib:EGM-14-001,bib:EGM-17-001}.

Jets are reconstructed from the PF candidates using the anti-\kt clustering algorithm \cite{bib:anti_kt} with a distance parameter of 0.4, as implemented in the \FASTJET library \cite{bib:fastjet}. Jets originating from {\cPqb} quarks are tagged with the \textsc{DeepCSV} algorithm \cite{bib:deep_csv}, which uses secondary vertices, together with other lifetime information, and exploits a deep neural-network architecture. For the chosen working point, the efficiency to select {\cPqb} quark jets is around 90\%, and the rate for incorrectly tagging jets originating from the hadronization of gluons or {\cPqu}, {\cPqd}, {\cPqs} quarks is about 10\%. Jet momentum is determined as the vectorial sum of all particle momenta in the jet and, based on simulation, is typically within 5 to 10\% of the true momentum over the whole {\pt} spectrum and detector acceptance. Additional $\Pp\Pp$ interactions within the same or nearby bunch crossings can contribute additional tracks and calorimetric energy depositions, increasing the apparent jet momentum. To mitigate this effect, tracks identified as originating from pileup vertices are discarded and an offset correction is applied to correct for remaining contributions. Jet energy corrections are derived from simulation studies so the average measured energy of jets becomes identical to that of particle-level jets. In situ measurements of the momentum balance in dijet, photon+jet, {\PZ}+jet, and multijet events are used to determine any residual differences between the jet energy scale in data and in simulation, and appropriate corrections are made \cite{bib:jet_energy_scale}. Additional selection criteria are applied to remove jets potentially originating from instrumental effects or reconstruction failures.

\section{Event selection}
The analysis strategy and selection criteria were designed before inspecting the signal region (SR) in the collision data. Events are chosen at trigger level with either a muon or an electron with \pt above 24 (27)\GeV for muons in the 2016 and 2018 (2017) samples, and 27 (32)\GeV for electrons in the 2016 (2017 and 2018) samples, following changes in the trigger reconstruction algorithms for these particles throughout the data-taking periods. The reconstructed isolated muon and electron candidates that match the trigger conditions must satisfy specific offline identification criteria, and are required to be in the region $\abs{\eta} < 2.4$, with $\pt>25\GeV$ (or ${>}28\GeV$ for 2017) for muons, and ${>}33\GeV$ (or ${>}30\GeV$ for 2016) for electrons. The muon and electron tracks should be within $\abs{d_z}<0.5\cm$ of the PV along the beam axis, and $\abs{d_{xy}}<0.2\cm$ in the transverse plane.

Events where multiple leptons satisfy the selection criteria are discarded to reduce the Drell--Yan background contribution. Pions are selected by searching for particle tracks with $\pt > 20 \GeV$ and charge opposite to that of the lepton. To reduce contamination from pion candidates misidentified as leptons, the requirements $\Delta\phi(\Pgm,\Pgp) > 0.09$ and $\Delta\phi(\Pe,\Pgp) > 0.05$ are imposed on the angle between the lepton and the candidate pion. These requirements remove the region where the lepton and the candidate pion overlap. Their effect on the signal efficiency is negligible. To ensure the compatibility of the pion track with the PV, we also require $\abs{d_z}<0.5\cm$ and $\abs{d_{xy}}<0.2\cm$. The pion with largest \pt that satisfies all these requirements is chosen for further analysis. Isolated photons with $\pt > 20 \GeV$ and $\abs{\eta} < 2.5$ are selected. The condition $\Delta\phi(\ell,\Pgg) > 0.04$ reduces the misidentification of the trigger lepton as the selected photon. The photon with the largest \pt that satisfies all these requirements is chosen for further analysis, and the event is retained if the pion-photon invariant mass is in the range $50 < \mpg < 100 \GeV$.

\subsection{Pion isolation}
\begin{figure}[!h]
\centering
\includegraphics[width=0.5\textwidth]{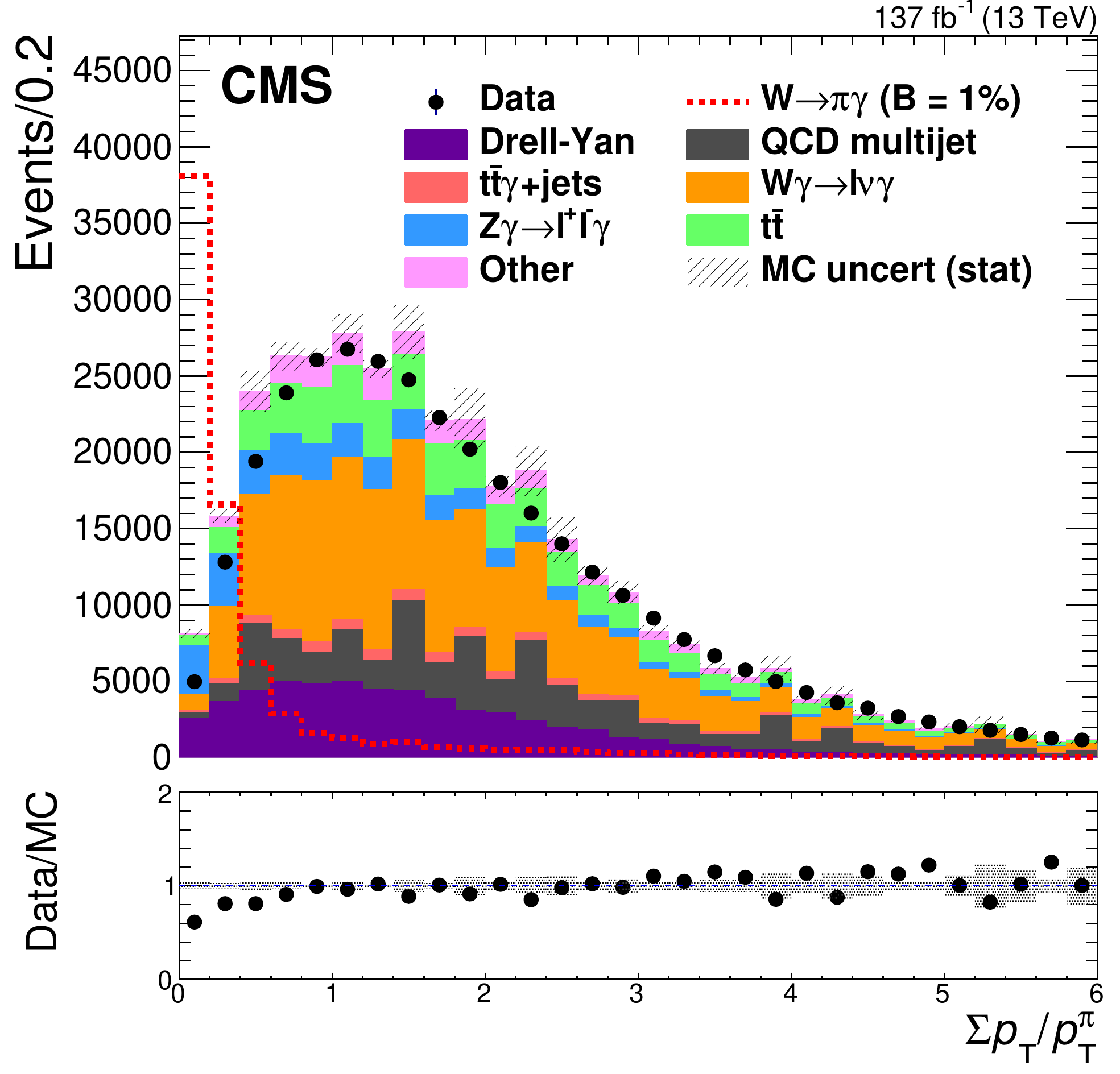}
\caption{Event distribution as a function of the pion-isolation variable $\Sigma\pt/\pt^{\Pgp}$. The simulated MC distribution for the signal is given by the dashed red line and corresponds to a 1\% branching fraction for the $\PWpm\to\Pgppm\Pgg$ decay. The statistical uncertainties in the data are small and thus not visible. In the lower plot, the ratio between data and the background component of the MC is shown. The gray bands represent the statistical uncertainty in the MC background.}
\label{fig:pion_isolation}
\end{figure}

Since the pion from the $\PWpm\to\Pgppm\Pgg$ decay originates from an electroweak process, minimal jet activity related to this decay is expected in its vicinity. For this reason, the analysis implements a pion-isolation variable based on the \pt sum of all the PF candidates (excluding the pion itself) contained in a cone with radius $\DR$ around the pion, divided by the pion $\pt^{\Pgp}$ ($\Sigma\pt/\pt^{\Pgp}$). The cone is defined to be $0.02 < \DR < 0.5$, with the lower bound helping to exclude from the sum the single charged particles generating two nearby reconstructed tracks. To reduce the contribution from pileup, charged particles entering the sum are required to originate from the PV, with the same $d_{xy}$ and $d_z$ criteria as used for the pion. Figure \ref{fig:pion_isolation} shows the distribution of the pion-isolation variable for signal events, which appear to be highly isolated, whereas background events indicate a clear presence of jet activity. A discrepancy between data and MC is observed for background events at low values of the pion-isolation variable. This discrepancy, independent of the data-taking period, does not affect the background characterization extracted from data, but somewhat reduces the effectiveness of the multivariate selection technique described in Section \ref{sec:MVA}, which is trained on simulated events. As far as the signal is concerned, this discrepancy is included in the systematic uncertainty as discussed in Section \ref{sec:systematics}.

\subsection{Multivariate selection}\label{sec:MVA}
\begin{figure}[!htb]
\centering
\includegraphics[width=0.49\textwidth]{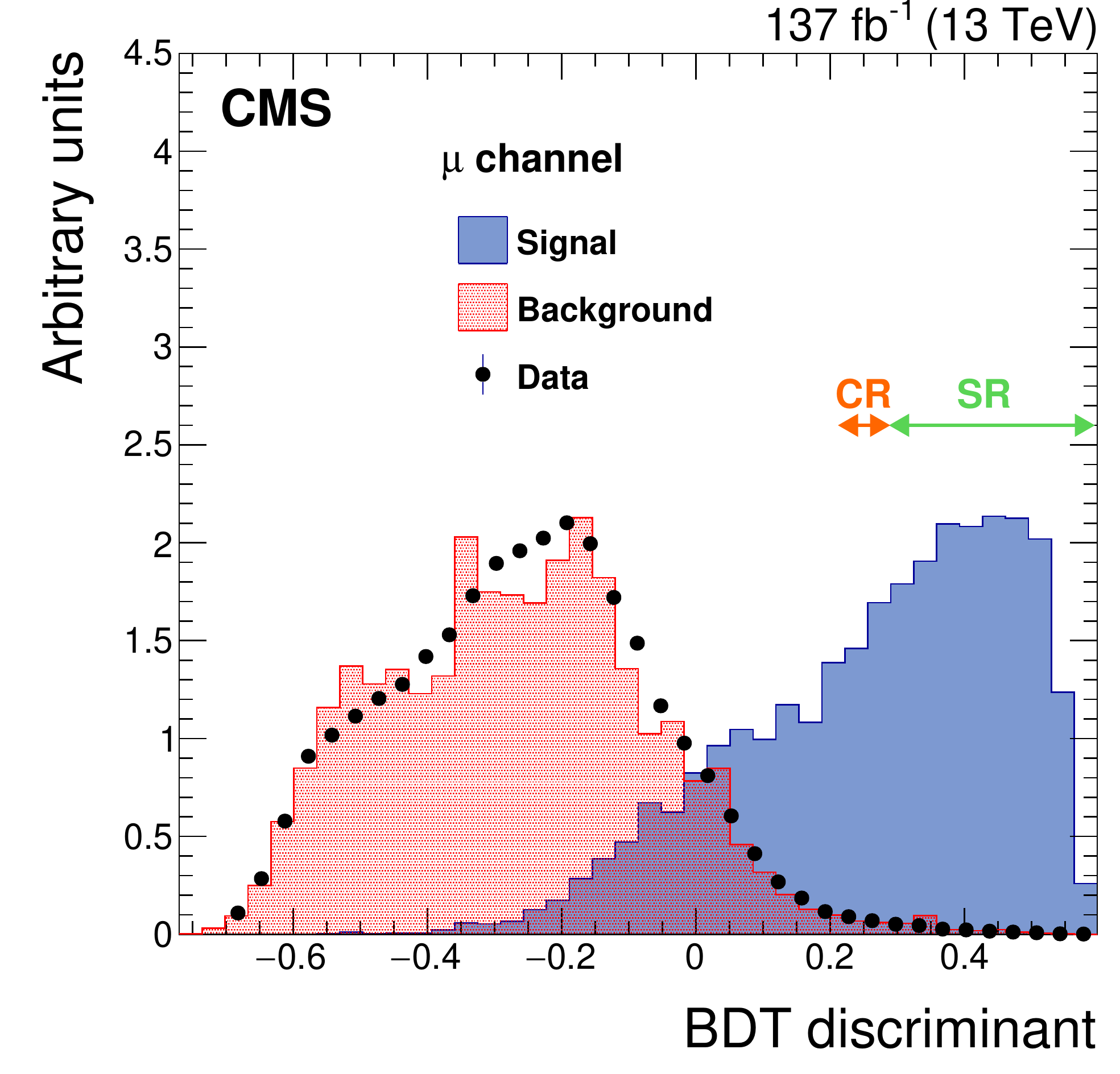}
\includegraphics[width=0.49\textwidth]{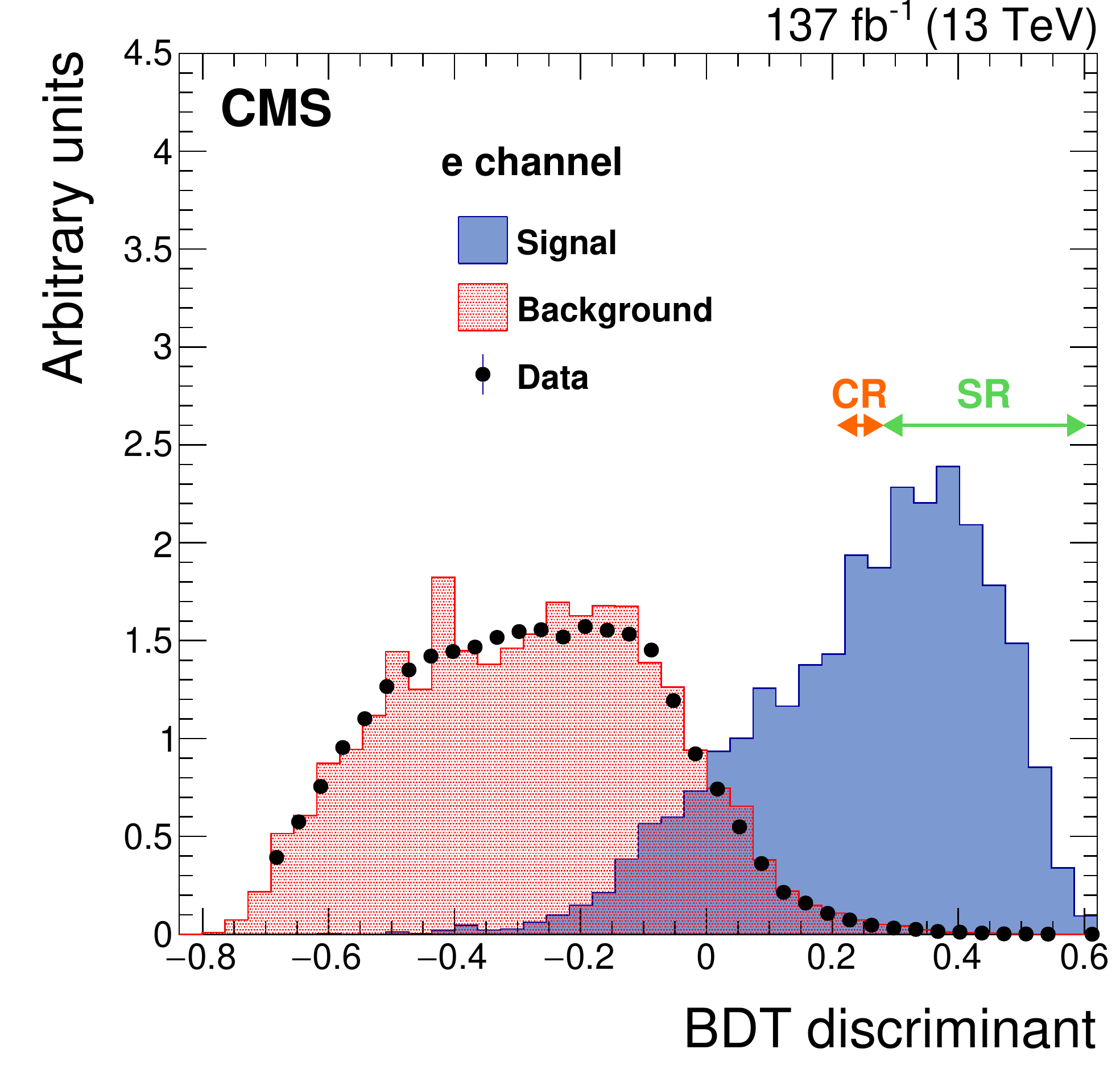}
\caption{Event distributions as a function of the BDT discriminant for the muon (\cmsLeft) and electron (\cmsRight) channels. The green and orange arrows indicate the intervals of the BDT discriminant used to define a signal (SR) and a control region (CR). The signal histogram is enhanced so that the signal and background histograms are normalized to the same area. The statistical uncertainties in the data are small and thus not visible.}
\label{fig:BDT_score}
\end{figure}

The final event selection is performed using a multivariate analysis technique. A boosted decision tree (BDT) classifier is trained using half of the signal and background MC events, and is validated through the other half. The requirement of a muon or an electron generates differences in the background distributions and the two cases are therefore handled as separate channels. The steps in the training and validation are carried out on the merged sample corresponding to the three years of data taking. The BDT input variables with good discriminating power and whose distributions are well described by the simulation are the transverse momenta of the lepton ($\pt^{\ell}$), pion ($\pt^{\Pgp}$), and photon ($\pt^{\Pgg}$), the \ptmiss in the event, the pion-isolation variable, and the number of the {\cPqb}-tagged jets in the event with $\pt > 25\GeV$ ($n_{\cPqb}$), which provides particularly good suppression of the background topologies characterized by no {\cPqb}-tagged jets. Distributions of the BDT discriminant for collision data with simulated signal and background are shown in Fig. \ref{fig:BDT_score}. The shape of the data is in agreement with the MC background.

Finally, thresholds for the BDT discriminant are defined for the two channels to identify a unique SR and a unique control region (CR) from the merging of the two lepton channels. These thresholds aim at obtaining the highest possible significance, while fulfilling the requirements discussed in the following. The CR, which includes a negligible signal contribution with respect to the SR, is chosen so that the $\mpg$ distribution contains enough events to be well fitted and to limit the impact of the systematic uncertainty in the background description. This is obtained for 0.206$<$ BDT discriminant $<$0.281 in the muon channel and 0.209$<$ BDT discriminant $<$0.269 in the electron channel. The SR should contain a number of events comparable to that in the CR, which is the case for values of the BDT discriminant $>$0.281 in the muon channel and $>$0.269 in the electron channel.

The products of the signal efficiency and the acceptance for the two channels, \eff, are computed separately for the 2016, 2017, and 2018 samples. They are evaluated as the ratio between the number of signal events in the MC sample in the SR, and the initial number of events generated in each channel. The efficiencies are given in Table \ref{tab:efficiencies} with their total uncertainties.

\begin{table}[!htb]
\centering
\topcaption{The product of signal efficiency and acceptance per year and per lepton channel.}
\begin{tabular}{ccc}
\hline
Year   & $\varepsilon_{\Pgm}$ & $\varepsilon_{\Pe}$\\
\hline
$2016$ & $0.12\pm0.01$       & $0.07\pm0.01$\\
$2017$ & $0.11\pm0.01$       & $0.07\pm0.01$\\
$2018$ & $0.12\pm0.01$       & $0.07\pm0.01$\\\hline                                                                             
\end{tabular}
\label{tab:efficiencies}
\end{table}

\section{Signal and background yield extraction}

The signal and background yields are extracted using an unbinned maximum likelihood fit to the $\mpg$ distribution in the SR. The probability density function (pdf) used to perform the fit is defined as follows:
\begin{linenomath}
\begin{equation}\label{eq:likelihood}
f = \left(\frac{N_{\text{sig}}}{N_{\text{sig}} + N_{\text{bkg}}(\mathrm{SR})} \, f_{\text{sig}} + \frac{N_{\text{bkg}}(\mathrm{SR})}{N_{\text{sig}} + N_{\text{bkg}}(\mathrm{SR})} \, f_{\text{bkg}}\right) \, \mathcal{G}, 
\end{equation}
\end{linenomath}
where $N_{\text{bkg}}(\mathrm{SR})$ is a floating parameter representing the number of background events in the SR, and $\mathcal{G}$ is a Gaussian pdf used to account for the nuisance parameters that model the systematic uncertainties. The functional form of the signal is determined fitting the $\mpg$ distribution obtained from simulation. The sum of a double crystal ball (DCB) function \cite{bib:DCB} and a Gaussian pdf is chosen to characterize the signal lineshape $f_{\text{sig}}$, whose parameters, excluding the normalization, are then fixed in the fit to the $\mpg$ distribution of the data.

The CR is used to estimate the functional form of the background directly from data. A linear polynomial ($f_{\text{bkg}}$) is chosen to describe the background. The values of the slope and the intercept (normalization) of this straight line can then vary in the fit to the $\mpg$ distribution of the data in the SR.

The number of signal events $N_{\text{sig}}$ is not sensitive to the contribution of {\PW} bosons that are not produced via \ttbar decay. It was verified in simulation that such processes are rejected by the requirement of \ttbar production with a large \pt lepton and jets originating from the hadronization of {\cPqb} quarks. Another contribution to the number of signal events that is not considered arises from the associated production of a top quark and a {\PW} boson, which is expected to be two orders of magnitude smaller than that from \ttbar events.

\section{Systematic uncertainties}\label{sec:systematics}
Systematic uncertainties are treated as nuisance parameters in the fit to the $\mpg$ distribution. These uncertainties are included as log-normal pdfs in the likelihood and will be discussed in the following in descending order of impact on the final result.

The \ttbar production cross section in $\Pp\Pp$ collisions at $\sqrt{s} = 13\TeV$, which is included in Eq. (\ref{eq:Nsig}), was measured by CMS using the 2015 data \cite{bib:ttbar_xsec} to be $815\pm43$\unit{pb}, where the total uncertainty is the quadratic sum of the statistical, systematic, and luminosity-related components. The integrated luminosities of the 2016, 2017, and 2018 data-taking periods are individually known with uncertainties in the 2.3--2.5\% range \cite{bib:cms_lumi_2016,bib:cms_lumi_2017,bib:cms_lumi_2018}, whereas the total 2016--2018 integrated luminosity has an uncertainty of 1.8\%; the improvement in precision reflects the uncorrelated time evolution of some systematic effects.

The choice of pdfs used to describe the background has an influence on the measured $\PWpm\to\Pgppm\Pgg$ branching fraction. To estimate this effect, a fit is performed to the data in the SR with the nominal background pdf, \ie, a linear function. The same distribution is then fitted again using an alternative pdf, namely an exponential. The following pull is calculated for the branching fraction:
\begin{linenomath}
\begin{equation}
\frac{\abs{\mathcal{B}_{\text{lin}} - \mathcal{B}_{\text{exp}}}} {\sigma_{\mathcal{B}_{\text{lin}}}},
\end{equation}
\end{linenomath}
where $\mathcal{B}_{\text{lin}}$ represents the value of the $\PWpm\to\Pgppm\Pgg$ branching fraction as extracted from the fit with a linear function, $\sigma_{\mathcal{B}_{\text{lin}}}$ is its uncertainty, and $\mathcal{B}_{\text{exp}}$ indicates the value of the branching fraction extracted from the fit using the exponential function. The pull amounts to 14.6\% and is taken as an estimate of the systematic uncertainty associated with the background parametrization.

The parameters representing the peak and width of the Gaussian component of the DCB pdf for the signal are extracted with their uncertainties through a fit to the simulated events. To include the limited event count in the MC sample, we take the width of the DCB function and its uncertainty as extracted from the fit to the MC events. Moreover, we include a 1\% uncertainty in the position of peak of the DCB function to account for the photon energy scale, which dominates the uncertainty in the pion-photon invariant mass. First, we fit the data in the SR fixing the values of the peak and the width of the DCB function to their central values (nominal case). We then repeat the fit after increasing and decreasing these parameters by their uncertainties (shifted parameters). We compare the results by calculating a set of pulls in the parameter of interest:
\begin{linenomath}
\begin{equation}
\frac{\abs{\mathcal{B}_{\text{nominal}}-\mathcal{B}_{\text{shifted}}}}{\sigma_{\mathcal{B}_{\text{nominal}}}},
\end{equation}
\end{linenomath}
where $\sigma_{\mathcal{B}_{\text{nominal}}}$ represents the uncertainty associated to $\mathcal{B}_{\text{nominal}}$. The largest pull of 10.6\% is an estimate of the systematic uncertainty associated with the parametrization of the signal. Since the other parameters of the signal lineshape are highly correlated with each other, as well as with the peak and the width of the Gaussian component of the DCB function, it is not necessary to float them within their uncertainties.

The contributions to the uncertainties in \eff include the statistical uncertainties, the BDT modeling, the modeling of kinematic variables for the signal in \PYTHIA, the scale factors that correct for differences between detector conditions in collision data and simulation, and the charge-misidentification in the tracking algorithm. To assess the uncertainty in modeling of the BDT, the performance of a BDT trained and validated with the nominal input variables is compared with that of a BDT trained with the same variables, and validated with variables to which a Gaussian spreading or a shift is applied. The width of the spreading corresponds to 5\% of the value of $\pt^{\ell}$, $\pt^{\Pgp}$, $\pt^{\Pgg}$, and \ptmiss. The pion isolation is shifted upwards by 10\%, based on the agreement between collision data and simulation observed in Fig. \ref{fig:pion_isolation}. The number of {\cPqb} jets, $n_{\cPqb}$, is increased or decreased by 1 for $5\%$ of the events in the training samples of 2016 and 2017, and for 10\% of the events in the training sample of 2018, thus accounting for the larger uncertainties in the jet energy calibration during part of the 2018 data taking. For energy and momentum, this spreading corresponds to the largest expected resolution. For charged particle tracks, it includes the uncertainties in the tracking efficiency. The largest change in \eff is $1\%$ for a given background rejection efficiency in the muon channel, and $2\%$ in the electron channel.

The procedure for modeling a process in \PYTHIA represents another source of systematic uncertainty, since inaccuracies of the model have an impact on the signal acceptance. As a first step in the evaluation of this uncertainty, we consider the \pt distribution of the generated signal {\PW} bosons, which is corrected to match that obtained from NLO \ttbar MC generated events in \POWHEG v2 (as discussed in Section \ref{sec:samples}). To assess the effect of a systematic change in the \ptW reweighting, the \ptW spectrum is shifted upwards and downwards by 5\%, which corresponds to the maximum uncertainty in the weights used. The performance of the BDT trained and validated through the nominal variables is compared to that of a BDT trained on the nominal variables, and tested using MC events where the \ptW-based reweighting is shifted by 5\%. The most significant changes in signal efficiency for a given background rejection efficiency are about 2\% and 3\% in the muon and the electron channels, respectively.

The signal acceptance also depends on the angular spectrum of the particles involved in the $\PWpm\to\Pgppm\Pgg$ decay. To evaluate the uncertainty on \eff related to the \PYTHIA modeling of the {\PW} polarization, we modify this angular spectrum. Specifically, we consider the generated distribution in $\theta^{*}$, the angle between the pion direction in the {\PW} boson rest frame and the direction of the {\PW} boson momentum in the laboratory. We use $\sin^{2}\theta^{*}$ and $\cos\theta^{*}$ distributions, corresponding to longitudinal and transverse polarizations of the {\PW} boson, respectively. By taking the ratio of each of these to the observed $\theta^{*}$ distributions, we obtain two sets of weights that are used to reweight the signal. In similarity with \ptW, a BDT is trained on a set of nominal events, and validated on a set that has the signal renormalization implemented. The most significant changes in terms of \eff for a given background rejection efficiency are about 3\% and 5\% for the muon and the electron channels, respectively.

An additional uncertainty of 1.4\% arises from the use of scale factors in the simulation, and an uncertainty of 1\% covers the effects of charge misidentification in the tracking algorithm. These uncertainties, derived from detector performance studies \cite{cit:cms_electron_reco,cit:cms_muonSyst,bib:EGM-14-001,bib:TRK-11-001}, are summed in quadrature to the statistical and to the other systematic uncertainties, described above, in the signal efficiency, as presented in Table \ref{tab:efficiencies}.

\section{Results}
\begin{figure}[htb]
\centering
\includegraphics[width=0.5\textwidth]{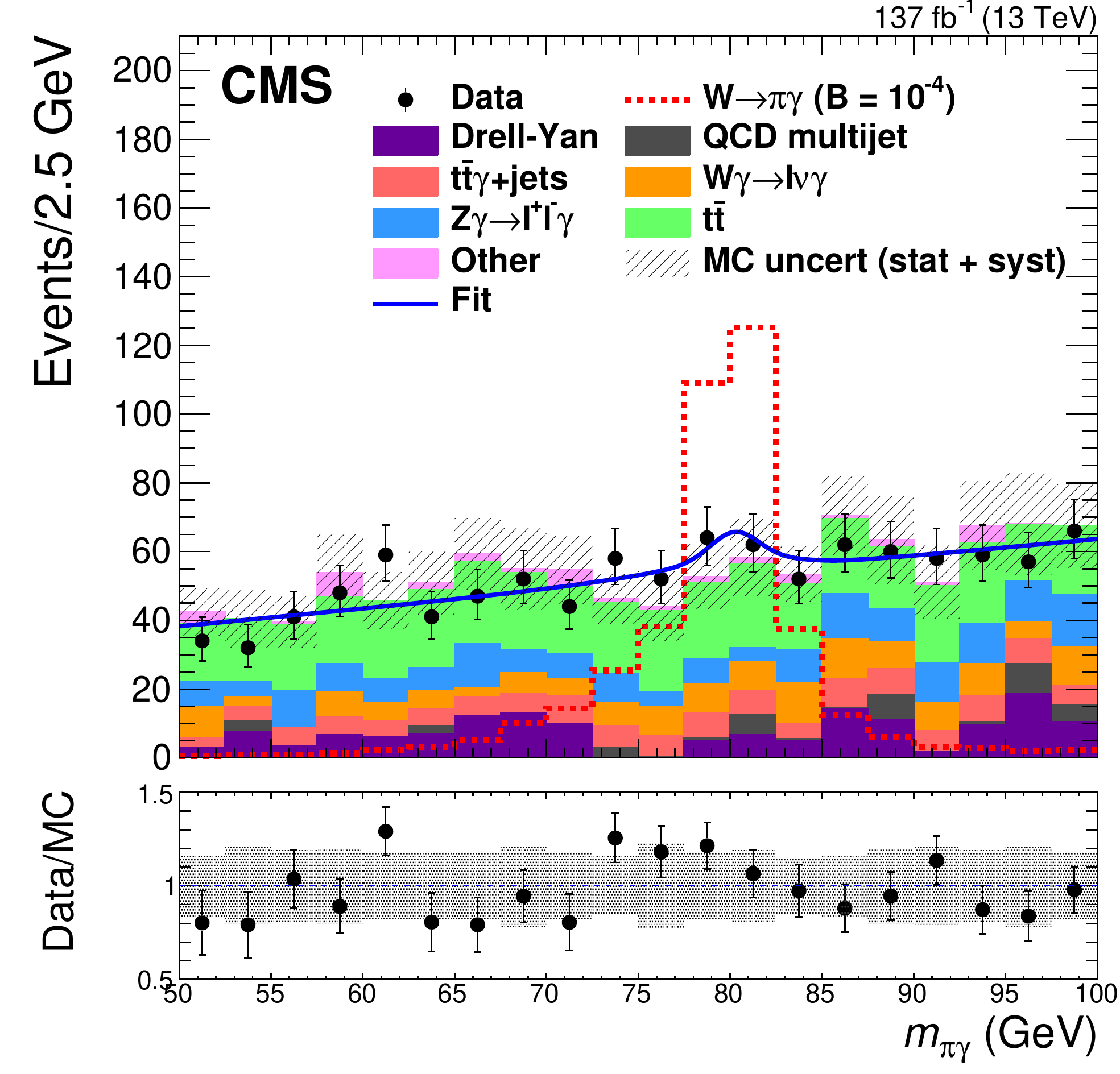}
\caption{Event distribution as a function of $\mpg$ for the combination of the lepton channels. The simulated MC distribution for the signal is given by the dashed red line and corresponds to a $10^{-4}$ branching fraction for the $\PWpm\to\Pgppm\Pgg$ decay. The uncertainties in the data are statistical only. The blue line represents the best fit to the data using the model described in Eq. (\ref{eq:likelihood}). In the lower plot, the ratio between data and the background component of the MC is shown. The gray bands represent the uncertainty (statistical + systematic) in the MC background.}
\label{fig:m_pi_gamma}
\end{figure}

The $\mpg$ distribution is shown in Fig. \ref{fig:m_pi_gamma} for the combination of the two lepton channels and the three data-taking periods.

No significant excess is observed above the expected background. An observed upper limit at 95\% \CL is set on the branching fraction of the {\PW} boson to a pion and a photon using the asymptotic \CLs method \cite{bib:CL_computation,bib:CLs}:
\begin{linenomath}
\begin{equation}
\mathcal{B}(\PWpm\to\Pgppm\Pgg) < 1.50 \times 10^{-5},
\end{equation}
\end{linenomath}
and the expected upper limit is $0.85^{+0.52}_{-0.29} \times 10^{-5}$.

The total uncertainty is dominated by the statistical contributions, which account for ${\approx}80\%$. The largest systematic uncertainties arise from the measurements of the \ttbar cross section (${\approx}5\%$) and the integrated luminosity (${\approx}2\%$).

\section{Summary}
A first search is reported for the rare decay $\PWpm\to\Pgppm\Pgg$ at the LHC. The search is based on the proton-proton collision data collected at a center-of-mass energy of 13\TeV at the CMS experiment in 2016--2018, corresponding to an integrated luminosity of 137\fbinv. Because of the high trigger thresholds for single photons, which make an inclusive search unsuitable at CMS, the measurement is performed using top quark-antiquark pair events, where one of the produced {\PW} bosons decays into leptons. This is the first search for $\PWpm\to\Pgppm\Pgg$ that adopts such a strategy. The rare decay is characterized by an isolated track, for which a specific pion-isolation variable is defined, and an isolated photon with large transverse momentum. The data are compatible with the background-only hypothesis. The upper limit on the branching fraction of the {\PW} boson to a pion and a photon is $1.50\times 10^{-5}$ at the confidence level of 95\%. 

Since this result is statistically limited, it can be significantly improved with larger integrated luminosities. Our study demonstrates the feasibility of a search for such rare decays of the {\PW} bosons at the LHC and defines a search strategy that can be adopted at future hadron colliders. This paves the way to increasingly precise measurements that could, on the one hand, provide effective validation of the QCD factorization formalism and, on the other, offer an alternative way to measure the mass of the {\PW} boson.

\begin{acknowledgments}
    We congratulate our colleagues in the CERN accelerator departments for the excellent performance of the LHC and thank the technical and administrative staffs at CERN and at other CMS institutes for their contributions to the success of the CMS effort. In addition, we gratefully acknowledge the computing centers and personnel of the Worldwide LHC Computing Grid for delivering so effectively the computing infrastructure essential to our analyses. Finally, we acknowledge the enduring support for the construction and operation of the LHC and the CMS detector provided by the following funding agencies: BMBWF and FWF (Austria); FNRS and FWO (Belgium); CNPq, CAPES, FAPERJ, FAPERGS, and FAPESP (Brazil); MES (Bulgaria); CERN; CAS, MoST, and NSFC (China); COLCIENCIAS (Colombia); MSES and CSF (Croatia); RIF (Cyprus); SENESCYT (Ecuador); MoER, ERC PUT and ERDF (Estonia); Academy of Finland, MEC, and HIP (Finland); CEA and CNRS/IN2P3 (France); BMBF, DFG, and HGF (Germany); GSRT (Greece); NKFIA (Hungary); DAE and DST (India); IPM (Iran); SFI (Ireland); INFN (Italy); MSIP and NRF (Republic of Korea); MES (Latvia); LAS (Lithuania); MOE and UM (Malaysia); BUAP, CINVESTAV, CONACYT, LNS, SEP, and UASLP-FAI (Mexico); MOS (Montenegro); MBIE (New Zealand); PAEC (Pakistan); MSHE and NSC (Poland); FCT (Portugal); JINR (Dubna); MON, RosAtom, RAS, RFBR, and NRC KI (Russia); MESTD (Serbia); SEIDI, CPAN, PCTI, and FEDER (Spain); MOSTR (Sri Lanka); Swiss Funding Agencies (Switzerland); MST (Taipei); ThEPCenter, IPST, STAR, and NSTDA (Thailand); TUBITAK and TAEK (Turkey); NASU (Ukraine); STFC (United Kingdom); DOE and NSF (USA).
     
    \hyphenation{Rachada-pisek} Individuals have received support from the Marie-Curie program and the European Research Council and Horizon 2020 Grant, contract Nos.\ 675440, 724704, 752730, and 765710 (European Union); the Leventis Foundation; the A.P.\ Sloan Foundation; the Alexander von Humboldt Foundation; the Belgian Federal Science Policy Office; the Fonds pour la Formation \`a la Recherche dans l'Industrie et dans l'Agriculture (FRIA-Belgium); the Agentschap voor Innovatie door Wetenschap en Technologie (IWT-Belgium); the F.R.S.-FNRS and FWO (Belgium) under the ``Excellence of Science -- EOS" -- be.h project n.\ 30820817; the Beijing Municipal Science \& Technology Commission, No. Z191100007219010; the Ministry of Education, Youth and Sports (MEYS) of the Czech Republic; the Deutsche Forschungsgemeinschaft (DFG) under Germany's Excellence Strategy -- EXC 2121 ``Quantum Universe" -- 390833306; the Lend\"ulet (``Momentum") Program and the J\'anos Bolyai Research Scholarship of the Hungarian Academy of Sciences, the New National Excellence Program \'UNKP, the NKFIA research grants 123842, 123959, 124845, 124850, 125105, 128713, 128786, and 129058 (Hungary); the Council of Science and Industrial Research, India; the HOMING PLUS program of the Foundation for Polish Science, cofinanced from European Union, Regional Development Fund, the Mobility Plus program of the Ministry of Science and Higher Education, the National Science Center (Poland), contracts Harmonia 2014/14/M/ST2/00428, Opus 2014/13/B/ST2/02543, 2014/15/B/ST2/03998, and 2015/19/B/ST2/02861, Sonata-bis 2012/07/E/ST2/01406; the National Priorities Research Program by Qatar National Research Fund; the Ministry of Science and Higher Education, project no. 02.a03.21.0005 (Russia); the Tomsk Polytechnic University Competitiveness Enhancement Program; the Programa Estatal de Fomento de la Investigaci{\'o}n Cient{\'i}fica y T{\'e}cnica de Excelencia Mar\'{\i}a de Maeztu, grant MDM-2015-0509 and the Programa Severo Ochoa del Principado de Asturias; the Thalis and Aristeia programs cofinanced by EU-ESF and the Greek NSRF; the Rachadapisek Sompot Fund for Postdoctoral Fellowship, Chulalongkorn University and the Chulalongkorn Academic into Its 2nd Century Project Advancement Project (Thailand); the Kavli Foundation; the Nvidia Corporation; the SuperMicro Corporation; the Welch Foundation, contract C-1845; and the Weston Havens Foundation (USA).\end{acknowledgments}

\bibliography{auto_generated}
\cleardoublepage \appendix\section{The CMS Collaboration \label{app:collab}}\begin{sloppypar}\hyphenpenalty=5000\widowpenalty=500\clubpenalty=5000\vskip\cmsinstskip
\textbf{Yerevan Physics Institute, Yerevan, Armenia}\\*[0pt]
A.M.~Sirunyan$^{\textrm{\dag}}$, A.~Tumasyan
\vskip\cmsinstskip
\textbf{Institut f\"{u}r Hochenergiephysik, Wien, Austria}\\*[0pt]
W.~Adam, T.~Bergauer, M.~Dragicevic, J.~Er\"{o}, A.~Escalante~Del~Valle, R.~Fr\"{u}hwirth\cmsAuthorMark{1}, M.~Jeitler\cmsAuthorMark{1}, N.~Krammer, L.~Lechner, D.~Liko, I.~Mikulec, F.M.~Pitters, N.~Rad, J.~Schieck\cmsAuthorMark{1}, R.~Sch\"{o}fbeck, M.~Spanring, S.~Templ, W.~Waltenberger, C.-E.~Wulz\cmsAuthorMark{1}, M.~Zarucki
% ...
\vskip\cmsinstskip
\dag: Deceased\\
1:  Also at Vienna University of Technology, Vienna, Austria\\
% ...
\end{sloppypar}
\end{document}